\documentclass[aps,twocolumn,superscriptaddress]{revtex4-2} 
\usepackage{graphicx}
\usepackage{dcolumn}
\usepackage{bm}
\usepackage[utf8]{inputenc}
\usepackage[T1]{fontenc}
\usepackage{mathptmx}
\usepackage{etoolbox}

\makeatletter
\def\@email#1#2{%
 \endgroup
 \patchcmd{\titleblock@produce}
  {\frontmatter@RRAPformat}
  {\frontmatter@RRAPformat{\produce@RRAP{*#1\href{mailto:#2}{#2}}}\frontmatter@RRAPformat}
  {}{}
}%
\makeatother

\usepackage{amssymb,amsmath,amsthm,amsfonts,amsbsy,mathrsfs}
\usepackage[english]{babel}
\usepackage{epsfig,amssymb,amsmath,amsthm,amsfonts,amsbsy,mathrsfs}
\usepackage{graphicx}
\usepackage{hyperref}
\usepackage{cleveref}

\usepackage{color}
\usepackage{bm}
\newcommand{\red}[1]{\textcolor{black}{#1}}

\begin{document}

\newcommand{\ntu}{Division of Physics and Applied Physics, School of Physical 
	and Mathematical Sciences, Nanyang Technological University, Singapore}
\newcommand{\create}{CNRS@CREATE LTD, 1 Create Way, \#08-01 CREATE Tower, Singapore 138602}
\newcommand{\cnr}{CNR--SPIN, Dipartimento di Scienze Fisiche, Universit\`a di 
	Napoli Federico II, I-80126, Napoli, Italy}

\title{Heterogeneous attenuation of sound waves in three-dimensional amorphous solids}

\author{Shivam Mahajan}
\affiliation{\ntu}
\author{Massimo Pica Ciamarra}
\email{massimo@ntu.edu.sg}
\affiliation{\ntu}
\affiliation{\cnr}
\affiliation{\create}

\date{\today}
%
\begin{abstract}
Sound waves are attenuated as they propagate in amorphous materials. 
We investigate the mechanism driving sound attenuation in the Rayleigh scattering regime by resolving the dynamics of an excited phonon in time and space via numerical simulations.
We find sound attenuation is spatiotemporal heterogeneous. 
It starts in localised regions, which identify soft regions within the material and correlate with low-frequency vibrational modes.
As time progresses, the regions where sound is primarily attenuated invade the system via an apparent diffusive process.
\end{abstract}
\maketitle

\section{Introduction}
The low-frequency vibrational excitations in crystals are plane waves (phonons). 
On the contrary, in amorphous solids, phonons attenuate~\cite{Masciovecchio2006, Baldi2010, monaco2009breakdown, Ruta2012, Baldi2013}, even in the harmonic regime~\cite{Mizuno2020}.
Sound attenuation originates from the nonaffine response~\cite{FlennerNonaffine, Baggioli2022} induced by amorphous solids' spatially fluctuating local elastic constants~\cite{SchirmacherPRL, SchirmacherEPL2006, Marruzzo2013a, Elliott1992, Graebner1986, Kawahara2011, kapteijns2021elastic}.
Some theories suggest accounting for the fluctuating elastic constants by describing an amorphous material as an elastic homogeneous medium punctuated by localized defects.
If this picture holds, the scattering by these localized defects drives sound attenuation, as in Rayleigh's model~\cite{Strutt1903}.

Theories suggesting that sound attenuation originates from localized defects \red{identified with the quasi-localised vibrational modes (QLMs)} successfully explain how the attenuation rate scales with the phonon wave vector and the stability of the material in the Rayleigh scattering regime~\cite{myPRL2}.
\red{However, they have difficulty in quantitatively relating the attenuation rate to QLMs' number and features~\cite{FlennerNonaffine}.}
One possibility is that the assumption that sound attenuation stems from localized defects is wrong, which is most probably the case in glasses that do not have a high degree of stability, and in two dimensions where low-frequency modes are not truly localized, the number of particles involved in these modes growing with the system size~\cite{Mizuno2017,kapteijns2018universal}.
Another possibility is that QLMs are just some of the scattering defects.
\red{In this respect, one has to consider that as the system size increases, the number of QLMs of a given frequency decreases as QLMs increasingly hybridize with plane waves.}


In this paper, we establish that localised regions playing the role of defects drive sound attenuation in three-dimensional amorphous solids and show that, in small systems, these defects relate to quasi-localised vibrational modes.
We investigate the nature of the attenuation process by resolving it in space and time via the introduction of a non-phononic displacement field measuring the deviation of the particles' displacement from the phononic one.
At short times, the non-phononic field is localized in a few regions, suggesting that attenuation is driven by defects.
As time advances, the non-phononic field gradually spreads via a diffusive process.
Mechanically soft regions may appear as anharmonic~\cite{Buchenau1992,  Ruffle2008} or harmonic quasi-localized soft vibrational modes~\cite{myPRL2} (QLMs)~\cite{Mizuno2014, kapteijns2018universal, richard2020universality, rainone2020pinching}, in finite-sized disorder systems. 
Conversely, in large systems, these soft regions are expected to be embedded within extended vibrational modes.
We show the regions where sound attenuation starts attenuating correlate with those where QLMs concentrate, and that QLMs strongly influence attenuation in poorly annealed glasses.
Overall, our results indicate that scattering by defects induces phonon attenuation in three-dimensional amorphous solids and suggest that these defects emerge as QLMs, in small and poorly annealed systems.

The paper is organized as follows.
We introduce our three-dimensional numerical model in Sec.~\ref{sec:model}, and review previous work investigating the dependence of the attenuation parameter on the phonons' frequency and the material properties in Sec.~\ref{sec:review}.
In Sec.~\ref{sec:attenuation}, we show that localized defects drive the attenuation of sound waves by investigating the spatiotemporal features of a non-phononic displacement field. Sec.\ref{sec:sound} and Sec.~\ref{sec:modecontribution} use diverse approaches to investigate how these defects correlate with quasi-localised vibrational modes.
In Sec.~\ref{sec:dynamics}, we show that the non-phononic field evolves via a diffusive process. Finally, we summarize our results and discuss future research directions in the conclusions.

\section{Numerical model and protocols \label{sec:model}}
\begin{figure}[t!]
 \centering
 \includegraphics[angle=0,width=0.48\textwidth]{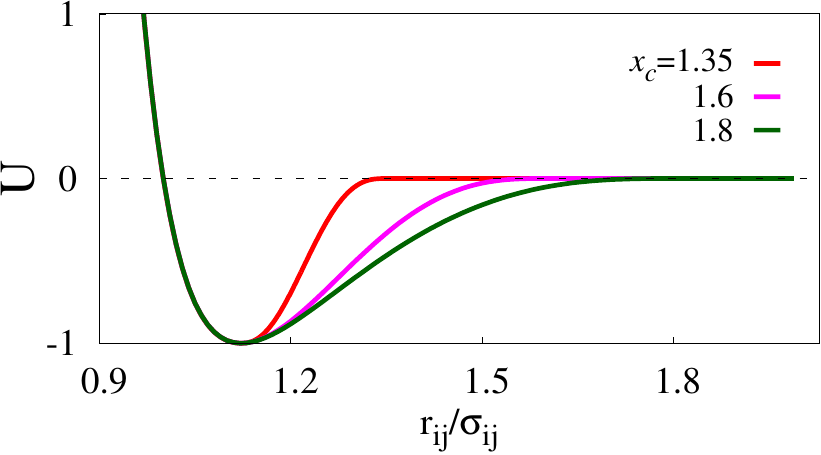}
 \caption{
 We consider a family of LJ-like interparticle potentials sharing the same repulsive part and different in the attractive well, which extends up to $r_{ij}/\sigma_{ij} = x_c$. 
  \label{fig:pot}
  }
\end{figure}

We investigate sound attenuation in model amorphous materials that differ in their degree of mechanical stability. 
Particles interact via LJ-like potentials $U(r,x_c)$ sharing the same repulsive part and differing in the extension of their attractive well~\cite{dauchotPotential},  which depends on the parameter $x_c$. 
Examples of potentials are shown in Fig.~\ref{fig:pot}. 
Previous works have shown that $x_c$ influences  the relaxation dynamics~\cite{chattoraj2020role} and the mechanical response~\cite{dauchotPotential, gonzalez2020mechanical, gonzalez2020mechanical2, Zheng2021}. 
In particular, the vibrational properties increasingly resemble those of stable glasses as the attraction range decreases. 

We follow the model of Ref.~\cite{chattoraj2020role} and consider polydisperse particles of diameter $\sigma_i$, drawn from a uniform random distribution in the range [0.8:1.2], in $d = 3$ spatial dimensions.
The potential has a repulsive and an attractive component.
The repulsive part of the potential follows the standard LJ functional form, 
\begin{equation}
U_r(r_{ij})=4\epsilon_{ij}\left[\left(\frac{\sigma_{ij}}{r_{ij}}\right)^{12}-\left(\frac{\sigma_{ij}}{r_{ij}}\right)^{6}\right],
\label{eq:repPot}
\end{equation}
where $\sigma_{ij}=(\sigma_i+\sigma_j)/2$, and acts for $r_{ij} \leq r_{ij}^{\rm min}=2^{1/6}\sigma_{ij}$.
The attractive part, which only acts for distances in the range
$2^{1/6}\sigma_{ij} \leq r_{ij} \leq x_c \sigma_{ij}$,
is given by
\begin{equation}
U_a(r_{ij})=\epsilon_{ij}\left[a_0\left(\frac{\sigma_{ij}}{r_{ij}}\right)^{12}\hspace{-0.4cm}-a_1\left(\frac{\sigma_{ij}}{r_{ij}}\right)^{6}\hspace{-0.2cm}+\sum_{l=0}^3{c_{2l}\left(\frac{r_{ij}}{\sigma_{ij}}\right)^{2l}}\right].
\label{eq:attPot}
\end{equation}
The parameters $a_0,a_1$ and $c_{2l}$ are chosen such that the potential $U(r_{ij})$ and its first two derivatives are continuous at the minimum $r_{ij}^{\rm min}$ and at the cutoff $r_{ij}^{(c)}=x_c \sigma_{ij}$, where the potential also vanishes.
In the following, we treat $m$, $\sigma$ and $\epsilon_{ij}=\epsilon$ as the units of mass, length, and energy.

We simulate systems with $N = 64000$ particles in a cubic box with periodic boundary conditions at fixed number density $\rho =1.07$.
We equilibrate the systems at temperature $4.0\epsilon$ well above the glass transition temperature for the considered models~\cite{chattoraj2020role} and then generate amorphous solid configurations by minimizing the energy via the conjugate gradient algorithm~\cite{FletcherConjGrad}.
In the following, we discuss the sound-damping and vibrational properties of these amorphous configurations.

\section{Dependence of sound attenuation on material properties~\label{sec:review}}
We investigate sound attenuation by exciting an acoustic wave excited at time $t=0$~\cite{Gelin2016, Mizuno2018} by giving each particle a velocity $\mathbf{v}_i^0=\mathbf{A}_T \cos(\boldsymbol{\kappa}{\bf r}_i^0)$, where $\bf{A} \boldsymbol{\kappa}=0$, considering $\boldsymbol{\kappa}$ in which two among $\kappa_x$, $\kappa_y$ and $\kappa_z$ are zero and $\kappa$ is the wave vector. 
We then evolve the system in the linear response regime where $\bold{\Dot{v}}_i(t)=-\sum_{j=1}^{N}{\mathcal{H}_{ij}.\bold{v}_j(t)}+\bold{\Dot{v}}_i^0\delta(t)$ with ${\mathcal H}_{ij}$ the Hessian matrix.
\begin{figure}[t!]
 \centering
 \includegraphics[angle=0,width=0.48\textwidth]{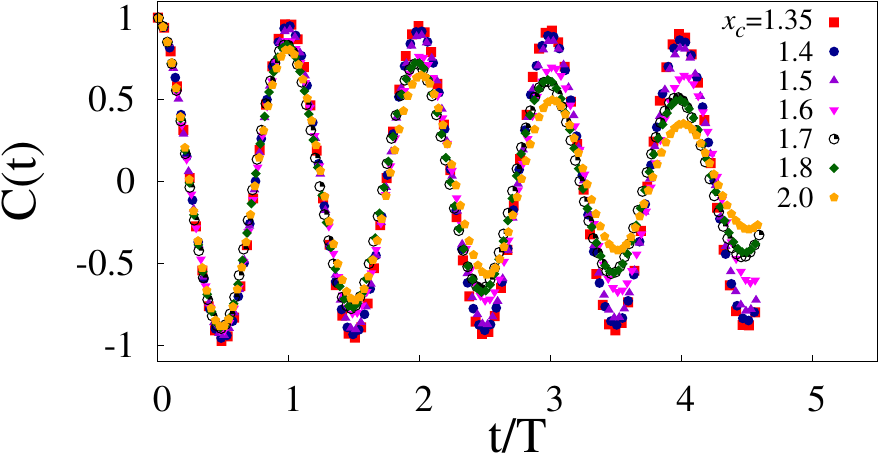}
 \caption{  
  Velocity autocorrelation function C(t) of a transverse phonon of wavevector $\kappa = \frac{2\pi}{L}(1,0,0)$ for different $x_c$, as a function of $t/T$, with $T$ the phonon's period. 
  \label{fig:vac}
  }
\end{figure}

The decay of the velocity auto-correlation function
\begin{equation}
C(t)=\frac{\sum_{i=1}^{N}{\mathbf{v}_i(0) \mathbf{v}_i(t)}}{\sum_{i=1}^{N}{\mathbf{v}_i(0) \mathbf{v}_i(0)}},
\label{eq:corrFn}
\end{equation}
illustrated in Fig.~\ref{fig:vac} demonstrates phonons' attenuation.
A fit of $C(t)$ to a damped exponential function $\cos(\omega t)e^{-\Gamma t/2}$ allows extracting the wave-vector dependence of the attenuation rate $\Gamma$ and frequency $\omega$ (or period $T = 2 \pi/\omega$).

\begin{figure}[t!]
 \centering
 \includegraphics[angle=0,width=0.48\textwidth]{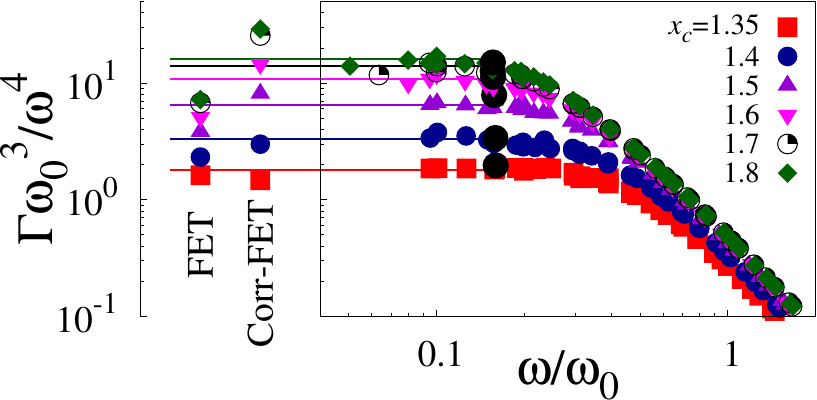}
 \caption{
Dependence of the scaled attenuation rate on the scaled frequency.
 The figure combines data for N = 32k, 64k, 256k, 512k, 2048k and 8192k, for which no size effects are apparent. 
 Symbols refer to different interaction potentials as in Fig.~\ref{fig:vac}.
 In the Rayleigh scattering regime, the correlated version of fluctuating elasticity theory accounts for the dependence of the attenuation rate on the material properties.
 In this manuscript, we resolve in space and time the evolution of low-frequency phonons with wave number $n=1$ in systems of N=64k particles, as indicated by the black circles and squares in the figure.
 Figure adapted from Ref.~\citenum{myPRL2}.
  \label{fig:rev}
  }
\end{figure}

In recent work~\cite{myPRL2}, we investigated how $\Gamma$ depends on the phonons' frequency, system sizes, and material properties for different cutoff $x_c$, focussing on the low-frequency scattering regime where $\Gamma \propto \omega^4$.
According to the theory of fluctuating elasticity~\cite{Maurer2004,SchirmacherPRL} (FET), which assumes that the local elastic properties are $\delta$-correlated in space, $\Gamma$ depends on the disorder parameter $\gamma$ regulating the dependence of the fluctuations of the shear modulus on the coarse-graining size $N$, $\sigma^2(N)/\mu^2 = \gamma/N$, $\mu$ being the average modulus. 
Specifically, FET predicts $\Gamma/\omega_0 \propto \gamma(\omega/\omega_0)^4$, with $\omega_0 = c_s/a_0$, $c_s$ the sound velocity of transverse waves and $a_0 = \rho^{-1/d}$ and $\rho$ the number density.
An extension of FET~\cite{Schirmacher2011}, which we termed correlated-fluctuating elasticity (corr-FET), considers the influence of the shear modulus' spatial correlation length, $\xi$.
If $\gamma \propto (\xi/a_0)^3$, as we observed~\cite{myPRL2} and expected if the disorder parameter is proportional to the correlation volume of the shear modulus, then corr-FET predicts $\Gamma/\omega_0 \propto \xi^3 \gamma(\omega/\omega_0)^4 \propto \gamma^2(\omega/\omega_0)^4$.
Our previous investigation, summarized in Fig.~\ref{fig:rev}, indicates sound attenuation is described by corr-FET rather than by FET, indicating the shear modulus' correlation length scale $\xi$ changes with $x_c$.
We found the same result holds for a different three-dimensional model.
These three-dimensional results are at variance with previous investigations of two-dimensional systems, where FET appears to hold~\cite{kapteijns2021elastic} and sound attenuation in the Rayleigh regime does not correlate with localized defects~\cite{Caroli2020}.

Interestingly, Rayleigh's~\cite{Strutt1903} original model, according to which the scattering by localized defects drives sound attenuation, reproduces corr-FET's prediction provided the defects satisfy some conditions.
While we gave indirect evidence suggesting that these conditions are met and that the scattering defects promoting sound attenuation are the quasi-localized vibrational modes, the relevance of the defect's picture remains controversial~\cite{FlennerNonaffine}.
We investigate this issue by resolving the damping of the phonons in time and space.

\section{Localized defects promote sound attenuation~\label{sec:attenuation}}
To unveil the microscopic mechanism driving sound attenuation, we spatially resolve the time evolution of the excited transverse phonons by introducing a non-phononic displacement field 
\begin{equation}
\mathbf{dr}_i^{\rm np}(t) = \mathbf{dr}_i(t)-\mathbf{dr}_i^{\rm ph}(t)
\end{equation}
where $\mathbf{dr}_i(t)$ is the displacement of particle $i$, and $\mathbf{dr}_i^{\rm ph}(t)$ the phononic displacement it would have in the absence of attenuation, 
\begin{equation}
    \mathbf{dr}_i^{\rm ph}(t) = -\frac{\mathbf{A}}{\omega} \sin(\mathbf{\kappa} \mathbf{r_i} - \omega t + \phi) + \frac{\mathbf{A}}{\omega} \sin(\mathbf{\kappa} \mathbf{r_i} + \phi),
\end{equation}
with ${\mathbf A}$ the polarization vector, ${\mathbf \kappa}$ the wavevector, $\omega = c_s\kappa$, $dt$ the integration timestep, $\phi$ a phase and $c_s$ the shear wave speed.

We filter out the influence of the slow-varying phononic field on the non-phononic displacement by studying it at stroboscopic times $sT$, with the $T$ the period.
We further compare $\mathbf{dr}^{\rm np}_i(sT)$ to its value at the relaxation time $s_\tau T$, with $s_\tau$ the smallest integer ensuring $s_\tau T = \tau = \leq 2\Gamma^{-1}$.
This comparison leads to the introduction of a local time-dependent sound attenuation parameter
\begin{equation}
    \Delta_i^2(sT) = \frac{\langle |\mathbf{dr}^{\rm np}_i(sT)|^2 \rangle}{\langle |\mathbf{dr}^{\rm np}_i(\tau)|^2 \rangle},
    \label{eq:AttPar}
\end{equation}
which is defined for $sT \leq \tau$.
Here, $\langle \cdot \rangle$ indicates an average are over phonons with different phases in $\phi \in [0,\pi/2]$ and the same $n^2=n_x^2+n_y^2+n_z^2$, where $\mathbf{\kappa} = \frac{2\pi}{L}(n_x,n_y,n_z)$.
\begin{figure}[t!]
 \centering
 \includegraphics[angle=0,width=0.5\textwidth]{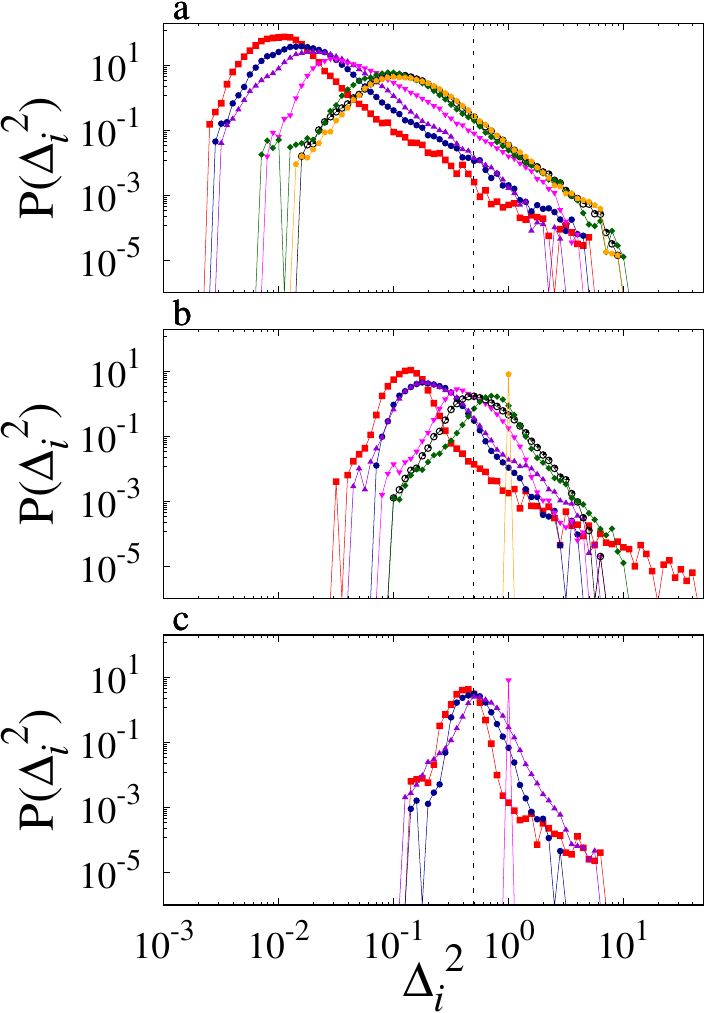}
 \caption{
Probability distribution of the local sound attenuation parameter $\Delta^2$ at times corresponding to one (a), four (b) and eight (c) phonon oscillation periods. 
Symbols identify different interaction potentials, as in Fig.~\ref{fig:vac}. 
We restrict the analysis to times $\leq 2\Gamma^{-1}$ and hence do not show data for large $x_c$ values in panel c.
The vertical dashed line marks the threshold $\Delta_c^2$ used to identify the particles with a large value of the attenuation parameter.
\label{fig:D2dist}
  }
\end{figure}

We have investigated the local sound attenuation parameter induced by different long-wavelength phonons.
We focus on the Rayleigh scattering regime by considering the longest-wavelength phonons. ($n^2=1$, circles in Fig.~\ref{fig:rev}). 
We average $\Delta_i^2(t)$ over $30$ phonons differing in their polarization or phase.

\begin{figure}[t!]
 \centering
\includegraphics[angle=0,width=0.5\textwidth]{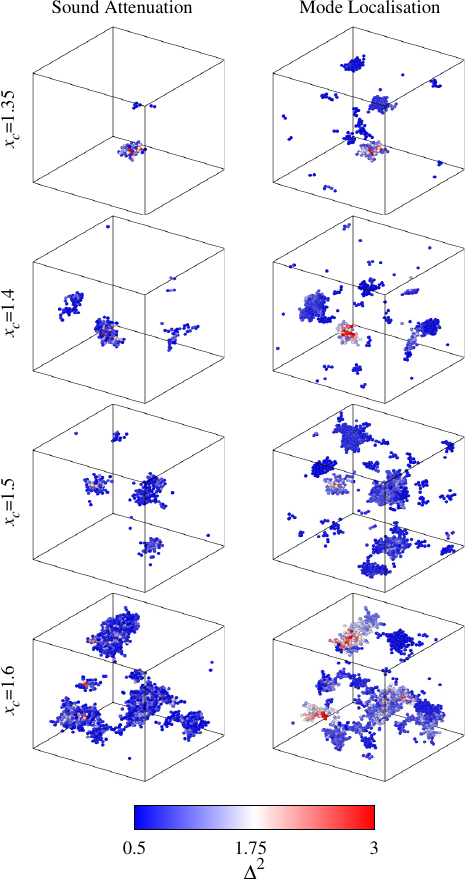}
 \caption{
The left column illustrates the particles with local phonon attenuation parameter $\Delta^2$ greater than 0.5, at a time corresponding to the period of the excited phonons. Data are averaged over phonons with $n^2 = 1$.
The right column illustrates the particles with localized mode participation (Eq.~\ref{eq:pi}) larger than $5\times10^{-5}$.
Different rows correspond to different interaction potentials, with the interaction range $x_c$ increasing from top to bottom.
Particles are colour-coded according to their $\Delta^2$ value.
The left column demonstrates that scattering defects drive sound attenuation in amorphous materials. The correspondence between the two columns demonstrates that these defects are the quasi-localized vibrational modes.
\label{fig:delta}
}
\end{figure}
Figure~\ref{fig:D2dist}a illustrates the probability distribution of $\Delta_i^2$ for the different interaction potentials at a time corresponding to one period of oscillation. 
The displacement is close to the expected phononic one for most particles, $\Delta_i^2 \ll 1$.
However, the distributions have extended right tails indicating the existence of a few particles whose motion strongly differs from the phononic one.
As time advances, the distribution's peak shifts towards $\Delta_i^2 = 1$, and the variance decreases, as apparent from Fig.~\ref{fig:D2dist}b,c that illustrate $P(\Delta_i^2)$ after four and eight oscillation periods.
At the relaxation time, the distribution becomes $\delta(\Delta_i^2-1)$. 
In the figure, this occurs in panel b for $x_c=2$ ($\tau(x_c=2) = 4T$), and in panel c for $x_c=1.6$ ($\tau(x_c=1.6) = 8T$).

The particles with local attenuation parameter $\Delta_i^2(t)$ larger than a threshold $\Delta_c^2$ identify the spatial region where sound attenuated.
The threshold $\Delta_c^2$ should be large enough to select as attenuated only the particles in the tail of the $P(\Delta_i^2)$ distribution, at early times.
It should also be smaller than $1$, the value $\Delta_i^2(t)$ at the relaxation time. 
We have checked that the choice of $\Delta_c^2$ is not critical as long as the above conditions are met and fixed in the following $\Delta_c^2=0.5$.

We illustrate in Fig.~\ref{fig:delta}(left column) the particles with $\Delta_i^2(t)> \Delta_c^2$ at a time corresponding to one phonon period.
Different rows correspond to different interaction potentials, increasing the attraction range from top to bottom.
The field is localized in compact clusters whose typical size increases with the attraction range, i.e., as the system becomes less stable.
Overall, this simple investigation unambiguously demonstrates that sound attenuation is promoted by localised regions that identify defects.

This observation poses two questions.
One question concerns the relation between the scattering defects and the quasi-localized vibrational modes~\cite{myPRL2}.
The second question concerns the growth dynamics of the non-phononic displacement field.
We address these two questions in Sec.~\ref{sec:sound} and Sec.~\ref{sec:dynamics}.

\section{Scattering defects and localized vibrational modes~\label{sec:sound}}
\begin{figure}[t!]
 \centering
 \includegraphics[angle=0,width=0.48\textwidth]{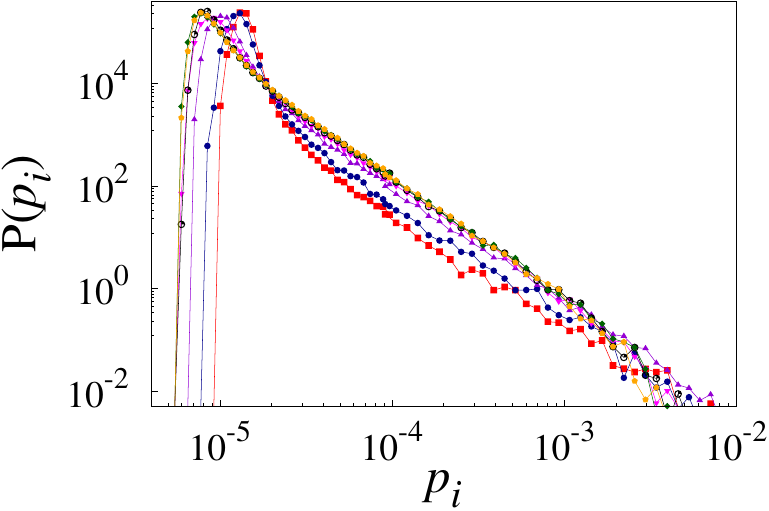}
 \caption{\label{fig:Pdistribution}
 Probability distribution of the individual particle mode participation $p_i^2$, for different $x_c$. 
 The fast decay ($\simeq 1/p_i^2$) reveals that only a small fraction of the particles have large displacements in the low-frequency vibrational modes.
   }.
\end{figure}

The soft localised regions driving phonon attenuation may be individually apparent as QLMs or hybridize with extended waves. How many soft regions emerge as QLM depends on the system size and, possibly, on the stability of the material.

To relate attenuation and vibrational modes we adopt a standard procedure to identify the particles involved in the QLMs~\cite{widmer2008irreversible}.
First, we determine the vibrational spectrum via the direct diagonalization of the Hessian matrix ${\mathcal H}_{ij}$ to obtain $k=1,2,...$ (normalized) eigenvectors ${\bf u}_k$ with associated eigenfrequencies $\omega_k$, $\omega_k \leq \omega_{k+1}$.
We average the squared displacement of particle $i$ in mode $k$ (${\bf u}_{k,i}^2$) over the lowest-frequency 0.015\% (30) eigenmodes~\cite{widmer2008irreversible}, to define an individual particle participation ratio,
\begin{equation}
    p_i = \langle  {\bf u}_{k,i}^2 \rangle.
    \label{eq:pi}
\end{equation}
If particle $i$ only participates in extended modes, then its $p_i$ value is of order $\mathcal{O}(1/N)$.
Particles with large $p_i$ are those primarily involved in the soft-localized modes.
We verified the number of considered modes is not critical, as $p_i$ is dominated by the contribution of the first few localized modes, with a frequency smaller than that of the Boson peak.

We illustrate the probability distribution $P(p_i)$ in Fig.~\ref{fig:Pdistribution}.
Regardless of the attractive range, the distribution peaks at small $p_i$ values of order 1/N, and decays approximately as $p_i^{-2}$ at large $p_i$.
The decay of $P(p_i)$ indicates that a few particles have unusually large displacements in the low-frequency modes. 
The amplitude of $P(p_i)$ at large $p_i$ values depends on $x_c$, as in short-ranged attractive potentials (small $x_c$) localized modes stiffen and shift to higher frequencies where they hybridize with extended ones~\cite{Edan_BP_QLM}.
\begin{figure}[t!]
 \centering
  \includegraphics[angle=0,width=0.48\textwidth]{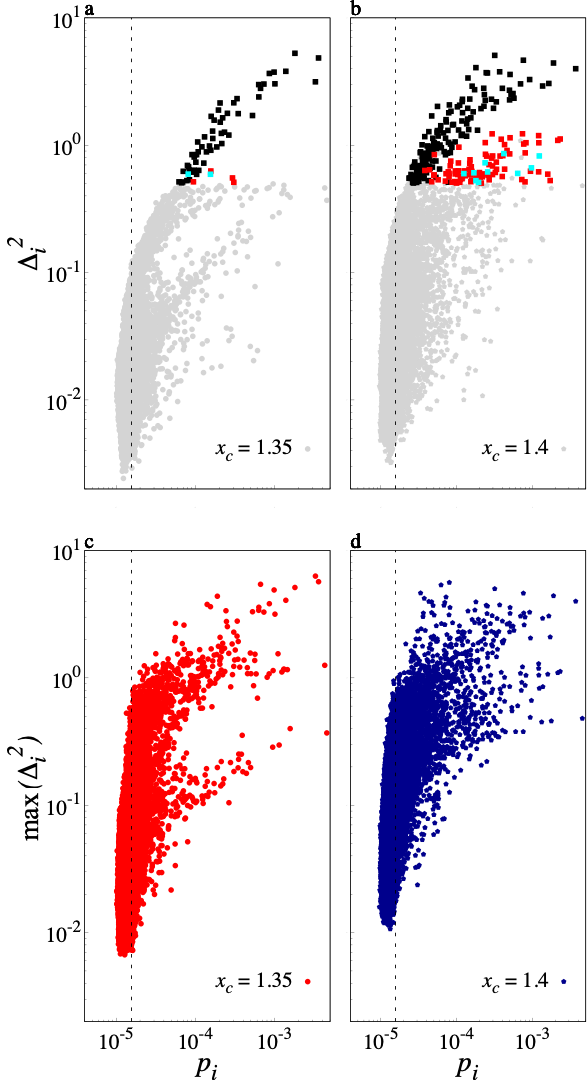}
 \caption{
Scatter plot of the particle attenuation parameter $\Delta_i^2$ against the individual participation ratio, $p_i$ for (a) $x_c = 1.35$, and (b) $x_c = 1.4$. $\Delta_i^2$ is evaluated after one period of oscillations, for $n=1$ phonons. 
\red{Blac, red and cyan point represent particles belonging to the first, second or third largest cluster illustrated in left columns of Fig.~\ref{fig:delta}.}
Panels (c) and (d) show corresponding plot obtained via associating to each particle the maximum of $\Delta_i^2$ over the modes with $n=1$ and $n=2$.
The dashed vertical lines mark the expected $p_i$ value ($1/N$) for particles only participating in extended modes.
  \label{fig:correlation}
  }
\end{figure}

Fig.~\ref{fig:delta}(right column) illustrates the location of the particles with $p_i > 5\times 10^{-5}$.
These particles are arranged into clusters that identify the spatial regions where QLMs localize.
The close correspondence between the left and the right columns of Fig.~\ref{fig:delta}, i.e., suggests QLMs play a key role in the attenuation process. 

To have a clearer insight into the influence of the modes on the attenuation process we investigate the $\Delta^2_i(t=T)-p_i$ scatter plot of Fig.~\ref{fig:correlation}a,b, where $\Delta_i^2$ is measured after exciting $n=1$ modes.
To interpret this plot, we consider that the particles most contributing to the attenuation process are those with a large $\Delta^2_i$.
These particles may occur in soft regions apparent as individual QLMs, in which they have large $p_i$, and in soft regions embedded in extended modes, in which case they have a small $p_i$. 
Since no particle with a small attenuation coefficient should occur in a soft region, no particle should have a small $\Delta^2_i$ and a large $p_i$.
This prediction is largely supported by Fig.~\ref{fig:correlation}a,b.
We do observe a few points representing particles with a somehow large value of the participation ratio and a small value of the attenuation parameter, e.g. $p_i\simeq 10^{-3}$ and $\Delta_i^2\simeq 10^{-2}$.
We have found these particles are not involved in the scattering of the considered $n=1$ phonons but of other ones. 
Indeed, by associating with each particle the maximum value of $\Delta_i^2$ over the modes with $n=1$ and $2$, we no longer observe particles with a small $\Delta_i^2$ and a large participation ratio, as in Fig.~\ref{fig:correlation}c,d. 
These findings echo previous results suggesting that, while quasi-localized modes influence the mechanical properties of amorphous materials, the relevant ones depend on the loading direction~\cite{Ding2014}.

Fig.~\ref{fig:correlation} demonstrates the absence of a one-to-one correspondence between $\Delta^2_i$ and $p_i$.
Rather, it suggests different branches depart from the cloud of points with small $p_i$ and $\Delta_i^2$ values.
We obtain insights into this organisation by performing a cluster analysis of the particles with $\Delta^2_i > \Delta_c^2$, i.e., of the particles visible in Fig.~\ref{fig:delta} (left column). 
Two particles are in the same cluster if in contact.
In Fig.~\ref{fig:correlation}a,b, we colour code in black, red and cyan the points representing particles belonging to the three largest clusters, and in grey all the others.
This study clarifies that the structures of Fig.~\ref{fig:correlation} reflect the particles' organisation into distinct clusters. 
Direct inspection further reveals that, within each cluster, $\Delta^2_i(t)$ is maximal in the centre and decays with the radial distance.

\section{Mode contribution\label{sec:modecontribution}}
We obtain further insights into the influence of the QLMs on sound attenuation via a normal modes analysis.
To this end, we project a phononic displacement field ${\bf r}$ on the vibrational modes orthonormal basis $\{{\bf u}_i\}$, ${\bf r} = \sum a_i {\bf u_i}$, $a_i={\bf r}\cdot{\bf u}_i$.
Exciting this phonon at a time $t=0$ by imposing appropriate velocities to the particles leads to a velocity field varying in time as $\dot{\bf p}(t) = \sum a_i \omega_i \cos(\omega_i t) {\bf u}_i$, where $\omega_i$ is the eigenfrequency of mode $i$.
The associated velocity autocorrelation function is
\begin{equation}
C(t) = \frac{\dot{\bf v}(t)\dot{\bf v}(0)}{\dot{\bf v}(0)\dot{\bf v}(0)}
= \frac{\sum a_i^2 \omega_i^2 \cos(\omega_i t)}{\sum a_i^2 \omega_i^2} \simeq \int P(a^2\omega^2) \cos(\omega t) d\omega
\label{eq:modes}
\end{equation}
Phonon damping with $C(t)=\cos(\omega_0 t) e^{-\frac{1}{2}\Gamma t}$ then occurs if the probability distribution of $a_i^2\omega_i^2$ is a Lorentzian, 
\begin{equation}
P(a^2\omega^2) = \frac{1}{\pi} \frac{\frac{1}{2} \Gamma^2}{(\omega-\omega_0^2)+(\frac{1}{2} \Gamma)^2 }
\end{equation}

\begin{figure}[t!]
\centering
\includegraphics[angle=0,width=0.35\textwidth]{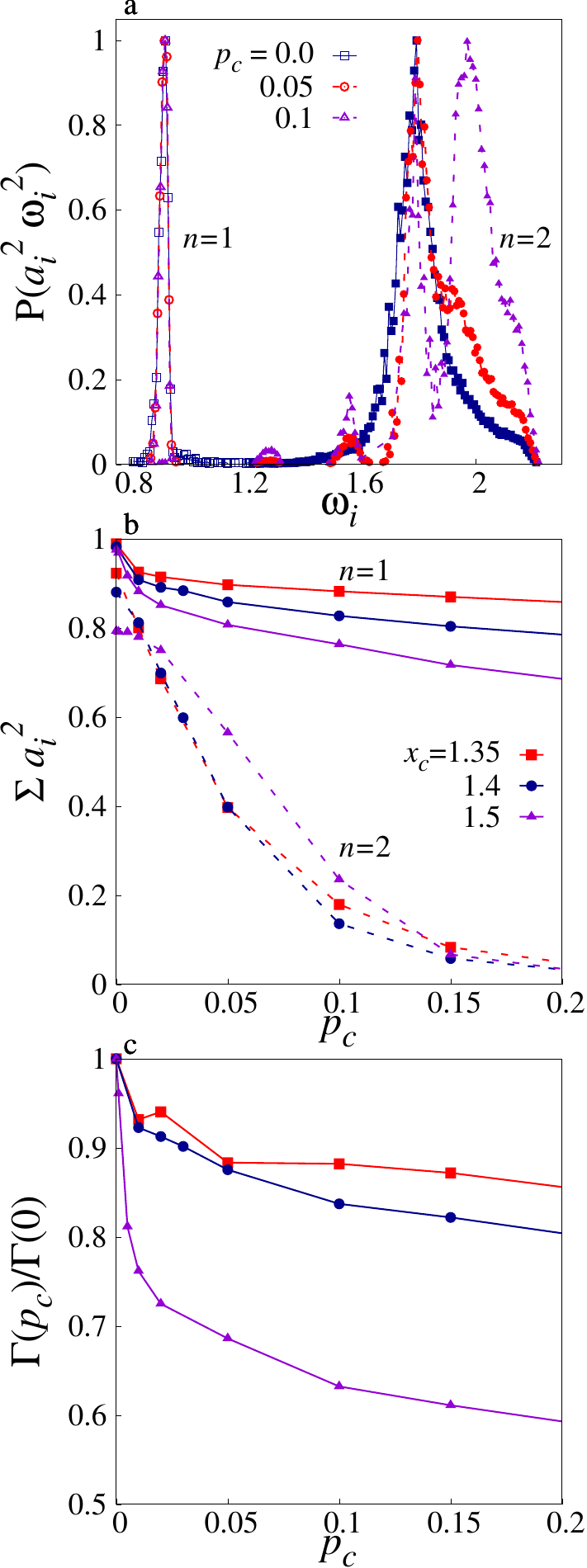}
\caption{
(a) Probability distribution $P(a^2 \omega^2)$ for $x_c=1.4$, for $n = 1,2$. The data is averaged over $100$ configurations, underlying $100$  phases and $d=3$ polarizations, considering all modes with participation ratios greater than $p_c=0,0.1,0.2$. 
(b) The sum of coefficients $a_i^2$ as a function of threshold $p_c$. 
(c) Ratio of $\Gamma_c/\Gamma_0$ as a function of $p_c$, where $\Gamma_c$ is the value at finite $p_c$ and $\Gamma_0$ represents $p_c=0$ (c).
\label{fig:projection}
}
\end{figure}

Using this projection formalism to investigate sound attenuation does not require the knowledge of all eigenmodes of the Hessian. 
Rather, one needs the modes contributing to the considered phonon, e.g., enough modes to ensure that $\sum a_i^2$ is approximately $1$.
Equivalently, one needs the modes with $a_i^2 > 1/dN$ that contribute more than random directions to the phonon, in $d$ spatial dimensions.
In general, more modes need to be considered as the phonon wavelength decreases.
In our $N=64000$ particle systems, we are able to get enough low-frequency modes to reasonably describe ($\sum a_i^2 > 0.96$) phonons with $n=1,2$, for all considered potentials.
As an example, we illustrate in Fig.~\ref{fig:projection}a numerical results for the distribution $P(a^2\omega^2)$ at $x_c = 1.4$.
We show results for $n=1$ (blue open squares) and $n=2$ (blue full squares).
At each $n$, the distribution is averaged over $100$ phases and $3$ polarizations, and $100$ different samples.
The Lorentzian-like shape of this distribution defines an attenuation coefficient, which is consistent with that estimated from the decay of the velocity autocorrelation function. 
Practically, to suppress noise, we obtain $\Gamma$ from a fit of the cumulative distribution to its expected functional form, $\frac{1}{2}+\frac{1}{\pi} \arctan(\frac{\omega-\omega_0}{\Gamma})$, focussing on a small range of frequencies close to $\omega_0$.

To assess the influence of soft modes on the attenuation coefficient, we restrict the above analysis to modes with a participation ratio larger than a threshold, $p_c$. 
Specifically, rather than describing the whole phonon by projecting it on the mode eigenbasis, we consider a fraction of it by projecting on the subspace spanned by the modes with participation ratio $p_i>p_c$.
In doing so, we capture a fraction $\sum_{p_i>p_c} a_i^2$ of the exited phonon, as we illustrate in Fig.~\ref{fig:projection}b, and modify the $P(a^2\omega^2)$ distribution.
As an example, Fig.~\ref{fig:projection}a shows how the original distributions (blue) change when considering only modes with $p_i \ge p_c = 0.05$ (red) and $\ge 0.1$ (magenta).
From this analysis, we are able to extract a $p_c$ dependent sound attenuation coefficient, $\Gamma(p_c)$, which is interpreted as the attenuation coefficient obtained after filtering out the influence of the modes with participation smaller than $p_c$.

We start by describing $n = 1$ phonons.
For $n = 1$, $\sum a_i^2$ drops gently as $p_c$ increases, as illustrated in Fig.~\ref{fig:projection}b, indicating that the phonons have a minimal projection on the localised modes. 
Nevertheless, the removal of this minimal projection influences the attenuation rate, as  in Fig.~\ref{fig:projection}c. 
The result is particularly apparent for $x_c=1.5$, whose vibrational properties share similarities with a poorly annealed glass~\cite{myPRL2}.
Henceforth, in this considered case, QLMs act as scattering sources at short times and sensibly influence the the attenuation coefficient.

QLMs' influence is more dramatic for $n=2$. 
In this case, $\sum a_i^2$ sensibly drops with $p_c$, as in Fig.~\ref{fig:projection}b,
and $P(a^2\omega^2)$ strongly depends on $p_c$, as in Fig.~\ref{fig:projection}a. 
These results imply the modes with a small participation ratio critically influence the modes with $n=2$ and their attenuation coefficient.
These modes influence $P(a^2\omega^2)$ so strongly that it can no longer be reasonably approximated by a Lorentzian. Henceforth, the $p_c$ dependence of the damping coefficient is ill-defined, which is why we do not illustrate $\Gamma(p_c)$ in Fig.~\ref{fig:projection}c.

\section{Sound attenuation spatiotemporal dynamics~\label{sec:dynamics}}
\begin{figure}[t!]
 \centering
 \includegraphics[angle=0,width=0.48\textwidth]{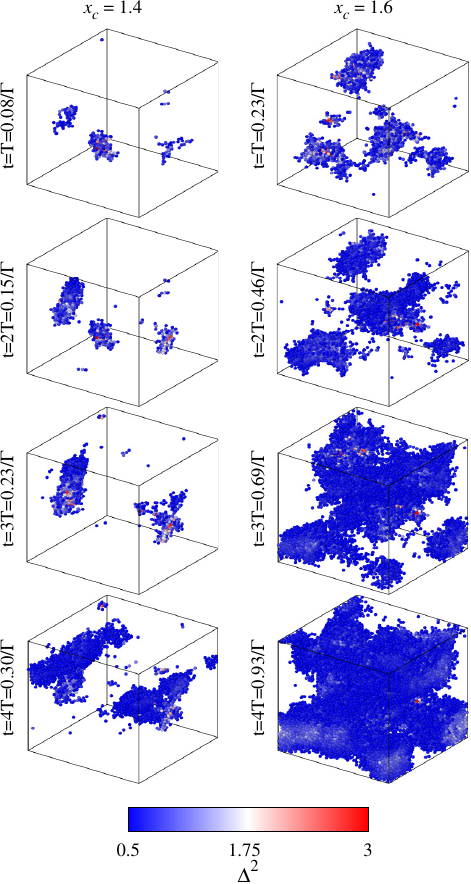}
 \caption{
  Particles with local phonon attenuation parameter $\Delta^2>0.5$ after 1,2,3 and 4 phonon oscillation periods for $x_c = 1.4$ (left) and $1.6$ (right).
  \label{fig:growth}
  }
\end{figure}
We investigate the spatiotemporal evolution of the local sound attenuation parameter in Fig.~\ref{fig:growth} by illustrating the particles with $\Delta_i^2 > \Delta_c^2$ at times corresponding to $1$, $2$, $3$ and $4$ oscillation periods, from top to bottom, for $x_c = 1.4$ (left column) and $x_c = 1.6$ (right).
The difference in the two columns reflects the interaction potential dependence of the attenuation parameter $\Gamma$.
Since $\Gamma(x_c=1.6) \simeq 3\Gamma(x_c=1.4)$, more particles have $\Delta_i^2>\Delta_c^2$ for $x_c=1.6$ than for $x_c=1.4$ after four oscillation periods. 
The figure reveals that attenuation proceeds through the growth of the clusters of attenuated particles observed at early times, rather than via the emergence of disconnected clusters.

\begin{figure}[t!]
\centering
\includegraphics[angle=0,width=0.48\textwidth]{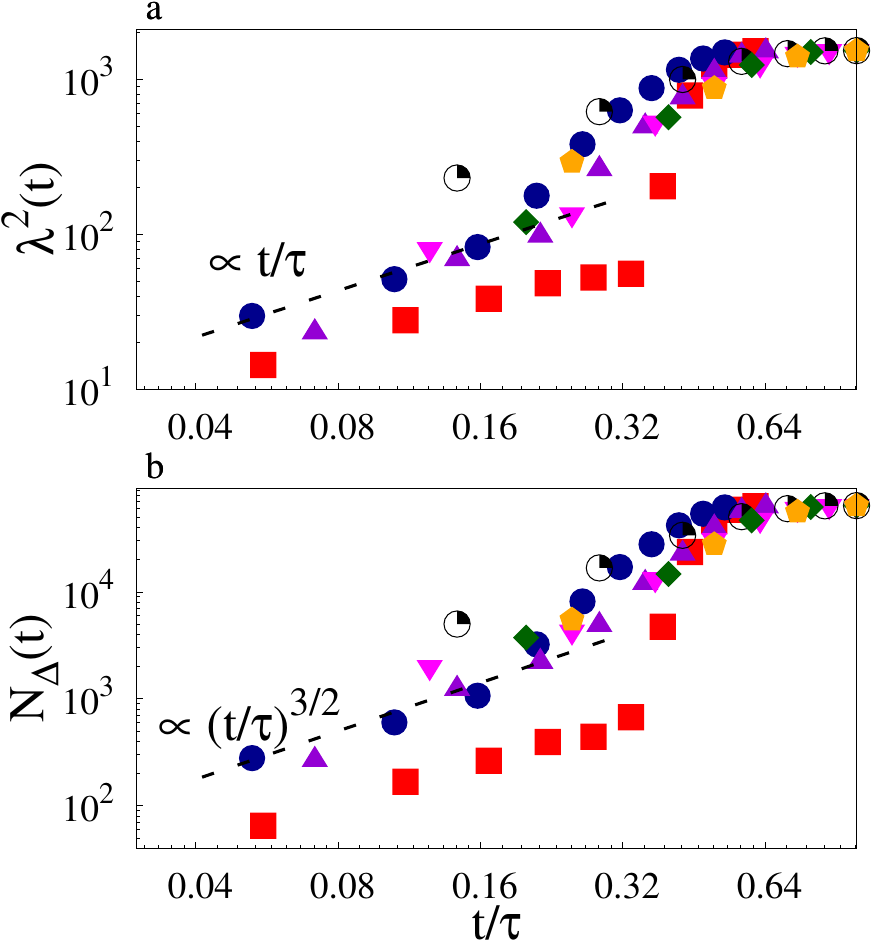}
\caption{
Time dependence of the squared linear size $\lambda^2$ of the largest cluster of particles with local damping parameter $\Delta^2_i(t) > 0.5$, and of the overall number of particles with $\Delta^2_i(t) > 0.5$. 
The early time growth of these quantities shows that the non-phononic field propagates via a diffusive process.
These quantities grow quickly at later times, as the peak of the $P(\Delta^2_i(t))$ distribution approaches the considered threshold value of $0.5$.
\label{fig:size}
}
\end{figure}

We quantify this growth process by studying the time dependence of the total number of particles with $\Delta_i^2 > \Delta_c^2$, $N_\Delta$, and of the linear size of the largest cluster, $\lambda=(N_L/\rho)^{1/3}$, with $N_L$ the number of particles in the cluster. 
Fig.~\ref{fig:size}a and b show that, at early times, $\lambda^2 \propto t/\tau$, and $N_\Delta \propto (t/\tau)^{3/2}$.
While attenuation in the harmonic regime we are investigating is deterministic as determined by the dephasing of the excited modes, Eq.~\ref{eq:modes}, these findings are compatible with a picture according to which the local attenuation parameter field $\Delta^2(\bf r)$ is localized in a few regions at very early times, and then spreads through a diffusive process.

At later times, both $\lambda^2$ and $N_\Delta$ grow sharply. 
This fast growth reflects the time dependence of the $P(\Delta_i^2)$ distribution, which we illustrated in Fig.~\ref{fig:Pdistribution}.
This distribution peaks at a $\Delta_i^2$ value that increases with time.
$\lambda$ and $N_\Delta$ grow sharply as the central peak approaches the threshold value $\Delta_c$ we used to identify the attenuated particles.
Indeed, we have observed the time of this fast growth depends on the chosen threshold.

\section{Conclusions} 
We have resolved the phonon attenuation in space and time, in the Rayleigh scattering regime, in three-dimensional model amorphous solids. 
To do this, we have introduced a non-phononic displacement field measuring the difference between the actual displacement of the particles and what it would be in the absence of attenuation. 
The spatiotemporal evolution of this field reveals that attenuation begins in specific localized defects and then spreads diffusively throughout the material.
These localized defects indicate softer areas within the system and often overlap with regions where low-frequency vibrational modes are concentrated.
Overall, these results support a picture according to which scattering by localised defects drives the attenuation of sound waves in three-dimensional amorphous materials.

Previous works suggesting that QLMs drive sound attenuation in three-dimensional amorphous solids capture the dependence of the attenuation parameter on material properties~\cite{myPRL2} but are quantitatively inaccurate~\cite{FlennerNonaffine}. 
Our results suggest this could occur as the defects inducing attenuation are not all apparent as individual QLMs. 
Rather, the fraction of defects apparent as QLMs is expected to decrease with the system size, due to their increased hybridization with extended waves. 
In large systems, other techniques might possibly be employed to identify localised soft regions~\cite{Richard2023}.
Relating sound attenuation to the properties of these regions is an interesting avenue of research that could benefit from previous approaches used to describe how the heterogeneities in the size of the grain boundaries affect sound attenuation in polycrystals~\cite{Arguelles2017}.

One of the most interesting future research avenues concerns the study of the dimensionality dependence of the attenuation process. 
In three dimensions, the few investigations of sound attenuation conducted so far suggest that the attenuation coefficient scales with material properties as predicted by correlated-fluctuating elasticity~\cite{Schirmacher2011}, which we termed correlated-fluctuating elasticity (corr-FET), $\Gamma/\omega_0 \propto \xi^3 \gamma(\omega/\omega_0)^4$, with $\xi$ the shear modulus' spatial correlation length.
Conversely, in two dimensions, previous results supported~\cite{Maurer2004,SchirmacherPRL} the FET scenario~\cite{SchirmacherEPL2006, Marruzzo2013a}, $\Gamma/\omega_0 \propto \gamma(\omega/\omega_0)^3$. 
In addition, in two dimensions, QLMs appear to be irrelevant~\cite{Caroli2020} as they strongly hybridize with plane waves~\cite{Bouchbinder2018} and disappear in large enough systems~\cite{Mizuno2017, Caroli2020}.
The two- and three-dimensional scenarios may be compatible if the correlation length $\xi$, which depends on the material properties in three dimensions, results weakly dependent on them in two dimensions, as we observed in a model system~\cite{Mahajan2022}, so that the corr-FET prediction reduces to the FET one. 
Alternatively, the defect picture may be relevant in three dimensions, not two. 
In this respect, it is worth noticing that size effects on sound attenuation are much stronger in two~\cite{Gelin2016} than in three dimensions ~\cite{Mizuno2017,Mizuno2018}.
This question might be addressed by investigating the non-phononic field in two-dimensional systems.

Our findings highlight a similarity between the behaviour of sound attenuation and the dynamical heterogeneities observed in supercooled liquids~\cite{Berthier2011a}.
In the case of phonon attenuation, we observed that particles with an attenuation parameter exceeding a certain threshold initially localize in specific regions, which gradually expand over time to encompass the entire system.
Likewise, in supercooled liquids, particles that move beyond a microscopic threshold during relaxation also localize in specific regions that grow over time and eventually encompass the entire system. 
Furthermore, as we have found attenuation correlates with QLMs, relaxation has been shown to correlate with QLMs ~\cite{widmer2008irreversible}. 
Since sound attenuation occurs in the linear response regime, this analogy indicates that dynamical heterogeneities may have a reversible component associated with particles' vibrations. It explains why they are not an optimal proxy of structural relaxation~\cite{Jack2014a, Li2022}.

\begin{acknowledgments}
We acknowledge support from the Singapore Ministry of Education through the Academic Research Fund Tier 2 (MOE-T2EP50221-0016) and Tier 1 (RG56/21), Singapore and are grateful to the National Supercomputing Centre (NSCC) of Singapore for providing the computational resources. We thank Darryl Seow Yang Han for fruitful discussions.
\end{acknowledgments}


%

\end{document}